\documentclass[12pt]{article}
\usepackage[dvips]{graphicx}
\begin{document}
\begin{center}
{\large\bf Resistance without resistors: An anomaly}\\
\vspace{0.50cm}
N. Kumar\\
Raman Research Institute, Bangalore 560080, India
\end{center}
\vspace{0.50cm}
\begin{abstract}
The elementary 2-terminal network consisting of a resistively ($R-$)
shunted inductance ($L$) in series with a capacitatively ($C-$) shunted
resistance ($R$) with $R = \sqrt{L/C}$, is known for its non-dispersive
dissipative response, $i.e.,$ with the input impedance $Z_0(\omega) = R$,
independent of the frequency ($\omega$). In this communication we examine
the properties of a novel equivalent network derived iteratively from this
2-terminal network by replacing everywhere the elemental resistive part
$R$ with the whole 2-terminal network. This replacement suggests a
recursion $Z_{n+1}(\omega) = f(Z_n(\omega))$, with the recursive function
$f(z) = (i\omega Lz/i\omega L + z) + (z/1+i\omega Cz)$.  The recursive map
has two fixed points -- an unstable fixed point $Z_u^\star = 0$, and a
stable fixed point $Z_s^\star = R$. Thus, resistances at the boundary
terminating the infinitely iterated network can now be made arbitrarily
small without changing the input impedance $Z_\infty (= R)$. This,
therefore, leads to realizing in the limit $n\rightarrow\infty$ an effectively
dissipative network comprising essentially non-dissipative reactive
elements ($L$ and $C$) only. Hence the oxymoron -- resistance without
resistors!  This is best viewed as a classical anomaly akin to the one
encountered in turbulence. Possible application as a formal decoherence
device -- the {\it fake channel} -- is briefly discussed for its quantum
analogue.\\

\end{abstract}
Consider an elementary 2-terminal $LCR$ network shown in Fig. 1. This
series -- parallel combination of the resistively ($R-$) shunted
inductance ($L$) in series with the capacitatively ($C-$) shunted
resistance ($R$) with $R = \sqrt{L/C}$, has a dispersionless dissipative
input impedance, $Z(\omega) = R$, independent of the circular frequency
($\omega$). This readily verifiable result is, of course, known, though
not as commonly as one would have expected it to be.  (The equivalence is
detailed in that, {\it e.g.}, the Nyquist-Johnson noise powers generated
by the two shunt resistors ($R$) at temperature $T$, say, combine to give
\setcounter{figure}{0}
\begin{figure*}
\includegraphics[width=5.0cm]{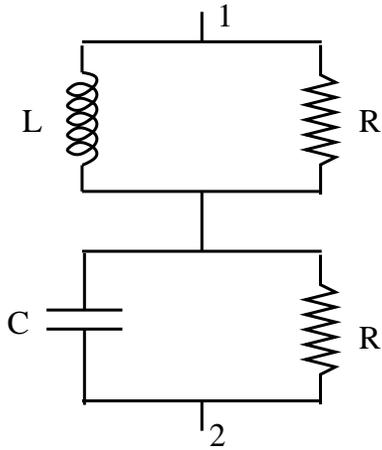}
\caption{Dispersionless 2-terminal LCR network with R = $\sqrt{L/C}$}
\end{figure*}
\begin{figure*}
\includegraphics[width=10.0cm]{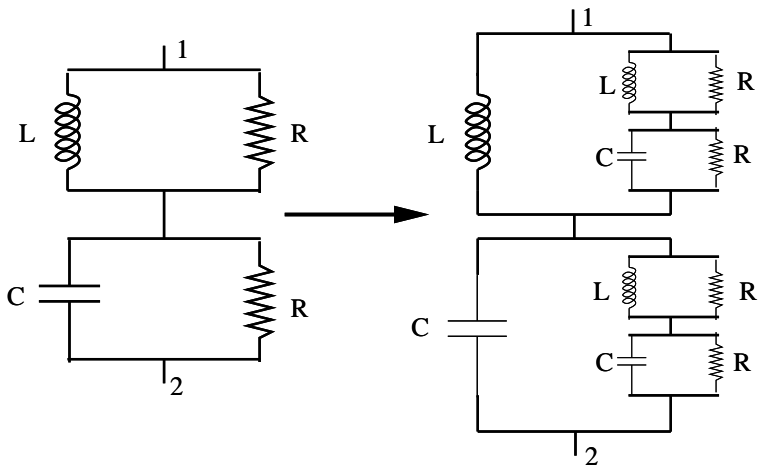}
\caption{Iteration of the 2-terminal LCR network with R = $\sqrt{L/C}$ replaced
by the whole 2-terminal network. Shown here is one stage of iteration}

\end{figure*}
a noise output at the (1-2)-terminal equal to that for a single resistance
$R$ at temperature $T$). The structure of this two-terminal network admits
iteration generating an equivalent network as indicated in Fig. 2.
Consider such an iterated network, but now terminated arbitrarily at the
boundary. With this, we can write the recursion relation
\begin{equation}
Z_{n+1} \equiv f(Z_n) = \frac{i\omega L Z_n}{i\omega L + Z_n} +
\frac{Z_n}{1+i\omega c Z_n}
\end{equation}

This recursion has two fixed points, $Z^\star = f(Z^\star)$ giving
$Z^\star = 0, R$. Linear stability analysis of these fixed points is
readily done. A perturbation $z_0$ about the fixed point $Z^\star = 0$,
iterates away giving $|z_{n+1}| = 2|z_n|$, making $Z^\star = 0$ an
unstable fixed point $Z_u^\star (=0)$.  Next, consider the fixed point
$Z^\star = R$.  A perturbation $z_0$ about $R$, iterates as
\[
z_{n+1} = \frac{1 - \omega^2 LC}{(1+i\omega \sqrt{LC})^2} z_n
\]
giving
\begin{equation}
\left| \frac{z_{n+1}}{z_n} \right|^2 = \left|\frac{1-\omega^2
LC}{1+\omega^2 LC} \right|^2 \leq 1.
\end{equation}
\begin{figure*}
\resizebox{1.0\columnwidth}{7 cm}{
\includegraphics{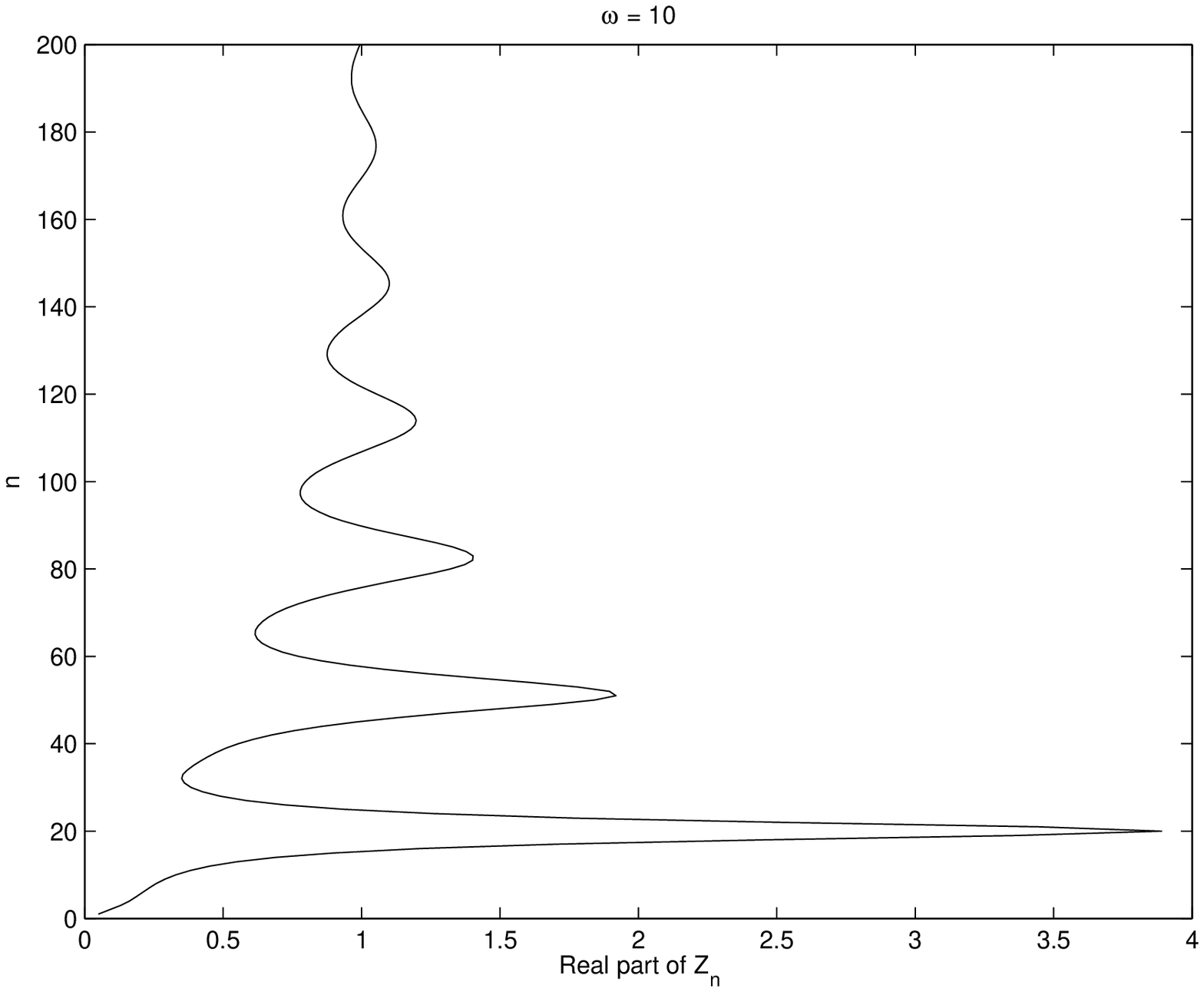}}
\caption{
Iteration of the 2-terminal network impedance ($Re
Z_n(\omega)$) initialized at $Z_0$ = 0.05 + $i$0.  Note the fast
convergence to the stable fixed point $Z_s^\star = 1$. Here $R = L = C =
1$,  and $\omega = 10$.}
\end{figure*}

\begin{figure*}
\resizebox{1.1\columnwidth}{7 cm}{
\includegraphics{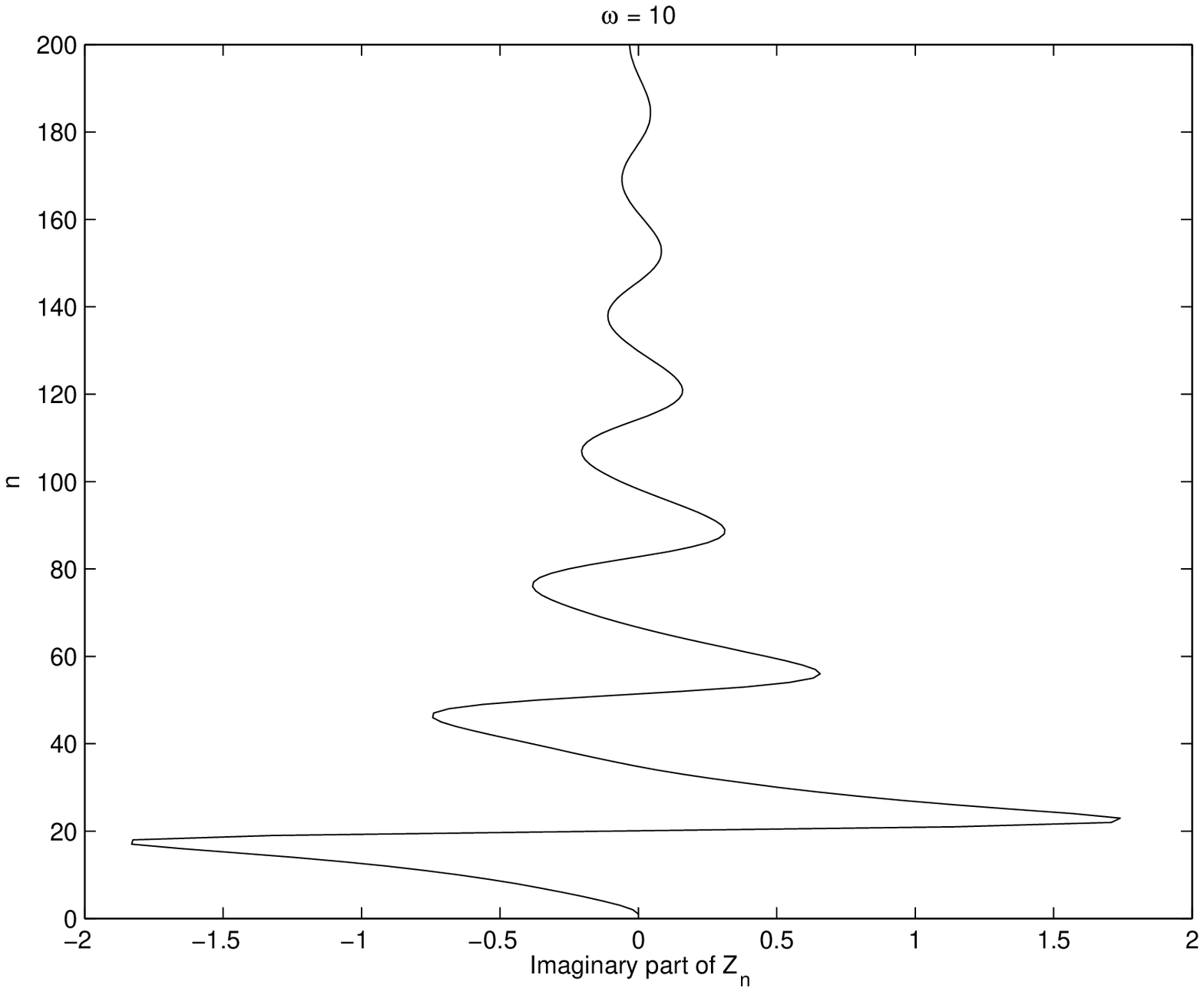}}
\caption{
Iteration of the 2-terminal network impedance $(Im
Z_n(\omega))$ initialized at $Z_0= 0.05 + i0$. Note the fast convergence
to the stable fixed point $Z_s^\star = 1$. Here $R = L = C = 1$,  and
$\omega = 10$.}
\end{figure*}

This makes the fixed point $Z_s^\star (= R)$ stable. The implication of
this fixed-point analysis is now straightforward. Terminating the network
at the boundary with $z_0 =  r_0 + ix_0$, where $r_0$ can be made
arbitrarily small (but non-zero positive), the impedance will iterate away
to the stable fixed point $Z_s^\star = R$ as $n \rightarrow\infty$.  This is,
however, so assuming that there are no other attractors. We have,
therefore, carried out the recursion in Eq. (1) numerically with different
initializations, and a typical evolution is shown in Figures 3 and 4.
Again, note the fast recursive convergence to the fixed point $Z_u^\star /
R\rightarrow1$.  {\it This is the all important point - an arbitrarily
small resistive termination at the boundary generates a finite resistance
$R = \sqrt{L/C}$ in the limit $n\rightarrow\infty$.} And this result suffices
for our purpose. (Inasmuch as the recursion holds for all values of the
frequency $\omega$, other attractors, if any, {\it e.g.,} a
period-doubling (2-cycle) attractor, would generate infinitely many
isospectral networks. Such attractors, or indeed a strange attractor,
should be interesting for network synthesis). The physical picture, of
course, is just this. The energy fed at the input terminal into the
infinitely iterated network appears to be absorbed effectively resistively
at the input terminal. But, in fact, it is really not dissipated there
instantaneously and locally -- it is cascaded away to the distant boundary
where it is ultimately dissipated. In a steady state {\it ac} response,
for instance, much energy remains stored in the reactive elements. This is
strongly reminiscent of what happens in fluid turbulence. There too,
energy fed at the large-scale eddy (integral, or energy regime) is cascaded away
progressively to smaller-scale eddies (inertial regime), and is ultimately
dissipated at the distant smallest (Kolmogorov) scale -- of viscosity. 
Indeed, the dissipation rate becomes independent of the viscosity in the
limit of vanishingly small viscosity! This is a classic example of the
classical dissipative {\it anomaly$^1$} -- the time-reversible symmetry
remains broken even as the symmetry-breaking parameter (the viscosity)
tends to zero, giving dissipation without dissipating elements!  The similarity to
our network is obvious (and not a little because of the inward bound
nature of our iterated network that makes the drawing in Fig. 2
increasingly more difficult beyond even the second stage of iteration). We
may note in passing that the iterated network is hierarchical in its
geometry.

Our analysis of the iterated network has implications for dissipative
quantum mechanics. It is known that there is no simple way of introducing
dissipation phenomenologically into a Hamiltonian quantum system without
inconsistencies$^2$.  A way out in the context of quantum transport has
been to introduce {\it fake channels$^{2,3}$}, such as transmission lines
that outcouple part of the wave amplitude causing the so-called stochastic
attenuation. Our infinitely iterated network is essentially a {\it
lumped-element} transmission line where the reactive elements can be
considered as part of the Hamiltonian system, and dissipation enters only
through the anomaly discussed above. A quantum analogue of our iterated
network would be the Cayley tree composed of 1-dimensional scatterers as
introduced by Shapiro$^4$ in the context of quantum conduction in parallel
resistors using splitters.  Further work along this line is in progress.

An interesting feature of our network is its invariance with respect to
certain correlated disorder, namely, that the condition $R = \sqrt{L/C}$
(fixed) allows us to vary $L$ and $C$ for a given $R$ at random with the
strong correlation, without leading to Anderson wave-localization$^{5,6}$
that would have blocked energy cascading. This is a case of purely gauge
disorder.

In conclusion, we have analyzed a 2-terminal {\it LCR} network which is
dispersionless and admits hierarchical iteration.  When infinitely
iterated, it gives an essentially reactive ($L$ and $C$) network and yet
provides dissipation - through an anomaly.  Possible application to
dissipative quantum systems is pointed out.  The network admits correlated
disorder without localization.

ACKNOWLEDGMENT. The author came to know of this dispersionless network
from V. Radhakrishnan (Rad), and would like to thank him for that. Its
provenance, however, remains untraceable. Thanks are also due to Andal
Narayanan for her help with the fixed-point analysis numerically.

\end{document}